\documentclass{aa}
\usepackage{psfig}
\usepackage{aabib99}

\begin{document}

\thesaurus{08 (02.12.1; 02.18.7; 03.13.4; 08.06.2; 09.13.2)}

\title{An accelerated Monte Carlo method to solve two-dimensional radiative
transfer and molecular excitation}
\subtitle{With applications to axisymmetric models of star formation}

\author{Michiel R. Hogerheijde\inst{1}
  \and Floris F. S. van der Tak\inst{2}}

\institute{Radio Astronomy Laboratory, University of California at
 Berkeley, Astronomy Department, 601 Campbell Hall \# 3411, Berkeley,
 CA 94720-3411, USA\\
 e-mail: michiel@astro.berkeley.edu
\and
 Sterrewacht Leiden, Postbus 9513, 2300 RA Leiden, The Netherlands\\
 e-mail: vdtak@strw.leidenuniv.nl}

\offprints{M. R. Hogerheijde}

\date{Received / Accepted}

\titlerunning{An accelerated Monte Carlo method}
\authorrunning{Hogerheijde \& van der Tak}

\maketitle


\begin{abstract}
  
  We present a numerical method and computer code to calculate the
  radiative transfer and excitation of molecular lines.  Formulating
  the Monte Carlo method from the viewpoint of cells rather than
  photons allows us to separate local and external contributions to
  the radiation field.  This separation is critical to accurate and
  fast performance at high optical depths ($\tau \ga 100$).  The
  random nature of the Monte Carlo method serves to verify the
  independence of the solution to the angular, spatial, and frequency
  sampling of the radiation field.  These features allow use of our
  method in a wide variety of astrophysical problems without specific
  adaptations: in any axially symmetric source model and for all atoms
  or molecules for which collisional rate coefficients are
  available. Continuum emission and absorption by dust is explicitly
  taken into account but scattering is neglected. We illustrate these
  features in calculations of (i) the HCO$^+$ $J$=1--0 and 3--2 emission
  from a flattened protostellar envelope with infall and rotation,
  (ii) the CO, HCO$^+$, CN and HCN emission from a protoplanetary disk
  and (iii) HCN emission from a high-mass young stellar object, where
  infrared pumping is important.  The program can be used for optical
  depths up to $10^3-10^4$, depending on source model.  We expect this
  program to be an important tool in analysing data from present and
  future infrared and (sub) millimetre telescopes.


\end{abstract}


\section{Introduction\label{s:intro}}

The dense and cool material in the interstellar medium of galaxies
plays an important role in the life cycle of stars, from the earliest
phases of star formation to the shells around evolved stars and the
gas and dust tori around active galactic nuclei. Line emission from
atoms and molecules, and continuum emission from dust particles, at
radio, (sub) millimetre and infrared wavelengths are indispensable
tools in the study of a wide variety of astrophysical problems. This
is illustrated by the large number of infrared and submillimetre
observatories planned for the near future, such as the Smithsonian
Millimeter Array (SMA), the Atacama Large Millimeter Array (ALMA), the
Far-Infrared and Submillimetre Space Telescope (FIRST) and the
Stratospheric Observatory for Infrared Astronomy (SOFIA). 

An essential step in the interpretation of the data from these
instruments is the comparison with predicted emission from models.
This paper presents a numerical method to solve the radiative transfer
and molecular excitation in spherically symmetric and cylindrically
symmetric source models.  At the comparatively low densities of
interstellar gas, the excitation of many molecules is out of local
thermodynamic equilibrium (LTE), and the transfer of line (and
continuum) radiation plays a significant role in determining the
molecular excitation \cite{leung:microt,black:iau197}. Geometry thus
becomes an important element, and the high spatial resolution of
current and future instruments often demands that at least
two-dimensional (axisymmetric) source structures are considered.  In
the implementation of our method discussed in this paper, we have
limited the source structure to spherical and cylindrical symmetries.
The large and often multidimensional parameter space further requires
a fast and reliable method, which needs to be easily applicable to
many different astrophysical problems.
 
This need for reliable and flexible tools calls for the use of Monte
Carlo techniques, where the directions of integration are chosen
randomly.  This approach was first explored by Bernes
\cite*{bernes:montec} for non-LTE molecular excitation; later, Choi et
al.~\cite*{choi:b335mc}, Juvela~\cite*{juvela:clumpy} and Park \&
Hong~\cite*{park:3dnonlte} augmented it and expanded it to multiple
dimensions. However, Monte Carlo methods can be quite slow, especially
at large optical depths ($\tau \ga 100$), which has limited their use
so far.  We will show that this problem can be overcome by using a
technique inspired on Accelerated Lambda Iteration: the local
radiation field and excitation are solved self-consistently and
separated from the overall radiative transfer problem (see \S
\ref{s:acceleration}). The greatest virtue of our code is its ability
to deal with a wide variety of source models for many atomic and
molecular species, with or without a dust continuum.  Although for any
individual problem a somewhat more efficient method could be
constructed (\S~\ref{s:noise}), the Monte Carlo approach frees the
user from having to fine-tune such a method and allows the user to
focus on the astrophysics of the problem at hand.

This paper does not discuss the influence of radiative transfer on the
source structure, through the thermal balance, ionization and
chemistry \cite[for
example]{takahashi:waterheating,ceccarelli:fir,doty:models}.  However, our
code is suited to become part of an iterative scheme to obtain
self-consistent solutions for the source structure including radiative
transfer and molecular excitation.

Throughout this paper, examples from studies of star formation will
serve to illustrate the various topics -- and to stress the link with
the analysis of observations. Sect.~\ref{s:simplemodel} introduces a
simple, spherically symmetric model of the core of an interstellar
cloud, collapsing to form a star. Sect.~\ref{s:method} then discusses
the coupled problem of radiative transfer and molecular excitation. It
introduces the two most commonly used solution methods, and shows that
these are closely related. This opens the possibility of constructing
a hybrid method which combines the benefits of both; the
implementation of this combined approach in our code is deferred to
the Appendix.  The paper continues by exploring the capabilities of
our code through a number of astrophysically relevant examples, based
on extensions of the simple one-dimensional model of
Sect.~\ref{s:simplemodel}. A brief summary concludes the paper in
Sect.~\ref{s:conclusion}.


\section{An illustrative, one-dimensional model\label{s:simplemodel}}

The formation of stars occurs in dense condensations within
interstellar molecular clouds, which collapse under the influence of
their own gravity. A widely used theoretical description of this
process, constructed by Shu \cite*{shu:selfsim}, starts with the
singular isothermal sphere,
\begin{equation}
\rho(r,t=0) = {{a^2}\over{2\pi G}} r^{-2}. \label{e:sis}
\end{equation}
Here, $\rho$ is the density as function of radius $r$ and time $t$,
$a$ is the isothermal sound speed, and $G$ is the gravitational
constant.

\begin{figure}[t]
 \begin{center}   
  \psfig{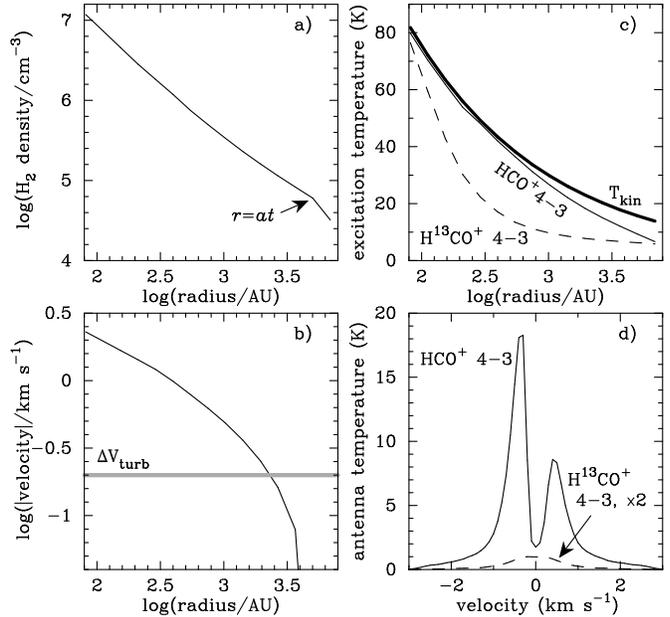}
\caption{Density (top left) and velocity (bottom left) structure of
the spherically-symmetric inside-out collapse model of Shu (1977) used
to illustrate the radiative transfer and molecular excitation problem
(\S \ref{s:simplemodel}). The excitation of HCO$^+$ (top right; solid
line) ranges from LTE in the dense, central regions to sub-thermal in
the lower density outer regions. Compared to the optically thin
excitation of H$^{13}$CO$^+$ (top right; dashed line), line trapping
significantly influences the HCO$^+$ excitation. The distribution of
the kinetic temperature is shown with the thick line for
comparison. The lower right panel shows the emergent HCO$^+$ and
H$^{13}$CO$^+$ $J$=4--3 line profiles in a $14''$ beam for a source at
140 pc. The asymmetric profile of the optically thick HCO$^+$ 4--3
line is characteristic of infall.
\label{f:shu77}}
\end{center}
\end{figure}

At $t=0$ collapse starts at the center ($r=0$). After a time $t$, all
regions $r<at$ are collapsing, with speed $v(r,t)$ increasing from 0
at $r=at$ to free-fall, $v\propto r^{-0.5}$, well within this
`collapse expansion wave' ($r\ll at$). Shu \cite*{shu:selfsim}
constructed a solution for the density and velocity field of the
collapsing core which is self-similar in the spatial coordinate
$x\equiv at$. The density follows a power-law behaviour as function of
radius, with $\rho \propto r^{-1.5}$ for $r\ll at$, $\rho \propto
r^{-1}$ just inside $r=at$, and the undisturbed $\rho \propto
r^{-2}$ outside $r=at$ (Fig.~\ref{f:shu77}). 

Many authors have tested this model against observations of cloud
cores and envelopes around young stellar objects (YSOs), e.g.\ Zhou et
al.\ \cite*{zhou:b335collapse}, Choi et al.\ \cite*{choi:b335mc},
Ward-Thompson et al.\ \cite*{wardthompson:n1333i2} and Hogerheijde \&
Sandell \cite*{mrh:scuba}.  Especially the spectral-line signature of
collapse (Fig.~\ref{f:shu77}d) has received much attention as a
probe of ongoing collapse, although this signature is shared by all
collapse models and is not unique to the particular model described
here. The exact line shape, however, depends quantitatively on the
adopted model.

The interpretation of this signature needs non-LTE radiative
transfer. Both collisional and radiative processes can excite
molecules, and for each transition a \emph{critical density} can be
defined where the two are of equal importance. At lower densities
radiation dominates, while at higher densities collisions drive the
level populations to thermodynamic equilibrium. The large range of
densities of star forming cores ensures that many molecules and
transitions will go through the entire range of excitation conditions,
while line emission will have a significant impact on the excitation
at the intensities and opacities expected for typical abundances of
many species, not only locally but throughout the envelope
(Fig.~\ref{f:shu77}c).

In the following we will use this model to illustrate our method of
solving the coupled problem of radiative transfer and excitation. In
particular, we will consider emission lines of HCO$^+$ and
H$^{13}$CO$^+$, which are readily observed and often used as tracers
of dense gas. The strong $J=1\to0$, $3\to2$ and $4\to3$ lines at
millimetre wavelengths have critical densities of $2\times 10^5$,
$4\times 10^6$, and $1\times 10^7$ cm$^{-3}$, using the molecular data
in Table~\ref{t:rates}.  We assume an abundance of HCO$^+$/H$_2=5
\times 10^{-9}$ and an isotopic ratio of 1:65 for
H$^{13}$CO$^+$:\,HCO$^+$.  The sound speed of the adopted model is
$a=0.24$ km~s$^{-1}$, its age $t=1\times 10^5$ yr, and its outer
radius 8000~AU. The total mass of the model is 0.73 M$_\odot$.  The
kinetic temperature follows $T_{\rm kin} = 30\,{\rm K}\, (r/1000\,{\rm
AU})^{-0.4}$, appropriate for a centrally heated envelope at a
luminosity of $\sim 2$ L$_\odot$ \cite[e.g.]{adams:ysosed}.
The turbulent line width of 0.2 km~s$^{-1}$ is smaller than the
systematic velocities except in the outermost part (Fig.~\ref{f:shu77}b).

\begin{table}
\caption{Molecular data used in this paper\label{t:rates}}
\begin{center}
\begin{tabular}{lrl}
\hline
Molecule & No. Levels & References \\
\hline
CO$^a$     &  6     & Green \& Thaddeus \cite*{green:corates} \\
CO$^b$     & 26     & Schinke et al.\ \cite*{schinke:co-h2} \\
HCO$^+$    & 21     & Monteiro \cite*{monteiro:hco+-h2} \\
CS         & 12     & Green \& Chapman \cite*{green:linearmols} \\
CN         & 15     & Black et al.\ \cite*{black:cn}$^c$ \\
HCN        & 36$^d$ & Green (1994, priv. comm.)$^e$ \\
o-H$_2$CO  & 20     & Green \cite*{green:h2co} \\
\hline
\end{tabular}
\end{center}

{\scriptsize a} Calculation presented in Appendix \ref{a:tests}.\\
{\scriptsize b} Calculation presented in \S\S~\ref{s:flat} and~\ref{s:disk}.\\
{\scriptsize c} Based on results of Green \& Chapman \cite*{green:linearmols} for CS.\\
{\scriptsize d} Levels up to $J=10$ in both the $\nu_2=0$ and
$\nu_2=1$ states.\\
{\scriptsize e} See http://www.giss.nasa.gov/data/mcrates.\\

\end{table}


\section{Solving radiative transfer and molecular excitation\label{s:method}}

\subsection{The coupled problem\label{s:radtrans}}

The equation of radiative transport reads, in the notation of Rybicki
\& Lightman \cite*{rybicki:radproc},
\begin{equation}
{{dI_\nu}\over{ds}}= -\alpha_\nu I_\nu + j_\nu, \label{e:di/ds1}
\end{equation}
or, equivalently,
\begin{equation}
{{dI_\nu}\over{d\tau_\nu}}= -I_\nu + S_\nu. \label{e:di/ds2}
\end{equation}
Here, $I_\nu$ is the intensity at frequency $\nu$ along a particular
line of sight parameterized by $ds$, $\alpha_\nu$ is the absorption
coefficient in units cm$^{-1}$, and $j_\nu$ the emission coefficient
with units erg~s$^{-1}$~cm$^{-3}$~Hz$^{-1}$~sr$^{-1}$. The second form
of the equation is a useful change of variables, with the source
function $S_\nu \equiv j_\nu/\alpha_\nu$ and the optical depth
$d\tau_\nu \equiv \alpha_\nu ds$. We consider both molecules and dust
particles as sources of emission and absorption ($j_\nu = j_\nu({\rm
dust}) + j_\nu({\rm gas})$; $\alpha_\nu= \alpha_\nu({\rm dust}) +
\alpha_\nu({\rm gas})$), but ignore scattering. Although not
impossible to include in our code, scattering effects are usually
negligible at wavelengths longer than mid-infrared.

When $\alpha_\nu$ and $j_\nu$ are known at each position in the
source, the distribution of the emission on the sky simply follows
from ray tracing. However, in many cases, $\alpha_\nu$ and $j_\nu$
will depend on the local mean intensity of the radiation field
\begin{equation}
J_\nu \equiv {1\over{4\pi}} \int I_\nu d\Omega.\label{e:jbar}
\end{equation}
Here, $J_\nu$ is the average intensity received from all solid angles
$d\Omega$, and $I_\nu$ is the solution of Eq.~(\ref{e:di/ds1}) along
each direction under consideration. The latter integration extends
formally to infinity, but in practice only to the edge of the source
with any incident isotropic radiation field like the cosmic microwave
background (CMB) as boundary condition.

For thermal continuum emission from dust, $j_\nu({\rm dust})$ and
$\alpha_\nu({\rm dust})$ are simply given by
\begin{equation}
j_\nu({\rm dust})= \alpha_\nu({\rm dust}) B_\nu(T_{\rm dust}),
\label{e:jnu_dust}
\end{equation}
where $B_\nu$ is the Planck function at the dust temperature $T_{\rm
  dust}$, and
\begin{equation}
\alpha_\nu({\rm dust}) = \kappa_\nu \rho_{\rm dust}, \label{e:alpha_dust}
\end{equation}
where $\kappa_\nu$ is the dust opacity in cm$^{-2}$ per unit (dust)
mass and $\rho_{\rm dust}$ is the mass density of dust.  Our code can
use any description of $\kappa_\nu$
\cite[e.g.]{ossenkopf:kappa,pollack:kappa,draine:andlee,mathis:mrn}.

In the case of emission and absorption in a spectral line,
$\alpha_\nu^{ul}({\rm gas})$ and $j_\nu^{ul}({\rm gas})$ are determined by
absorption and emission between radiatively coupled levels $u$ and $l$
with populations (in cm$^{-3}$) $n_u$ and $n_l$. The energy difference
between levels $\Delta E = E_u - E_l$ corresponds to the rest
frequency of the transition, $\nu_0 = \Delta E / h$, where $h$ is
Planck's constant. The emission and absorption coefficients between
levels $u$ and $l$ are strongly peaked around $\nu_0$ with a frequency
dependence described by a line-profile function $\phi(\nu)$,
\begin{eqnarray}
j_{\nu}^{ul}({\rm gas})      & = & 
  {{h\nu_0}\over{4\pi}} n_u A_{ul} \phi(\nu),\label{e:jnu}\\
\alpha_{\nu}^{ul}({\rm gas}) & = & 
  {{h\nu_0}\over{4\pi}} (n_l B_{lu} - n_u B_{ul})\phi(\nu) \label{e:alpha}.
\end{eqnarray}
The Einstein $A_{ul}$, $B_{lu}$, and $B_{ul}$ coefficients determine
the transition probabilities for spontaneous emission, absorption, and
stimulated emission, respectively, and depend on molecule. In
most interstellar clouds the line profile is dominated by Doppler
broadening due to the turbulent velocity field
\begin{equation}
\phi(\nu)= {c\over{b\nu_0\sqrt{\pi}}}
\exp \left( - {{c^2 (\nu - \nu_0)^2}\over{\nu_0^2 b^2}} \right )\label{e:phi},
\end{equation}
where the turbulence is assumed to be Gaussian with a full width $b$.
In the presence of a systematic velocity field, the line profile is
angle-dependent and the projection of the local velocity vector onto
the photon propagation direction enters $(\nu - \nu_0)$.

Together, collisions and radiation determine the level populations through
the equation of statistical equilibrium,
\begin{equation}
\begin{array}{l}
n_l \left[ \sum_{k<l}\, A_{lk} + 
\sum_{k\neq l}\, (B_{lk} J_\nu + C_{lk})\right] =  \\[0.3cm]
 \sum_{k>l}\, n_k A_{kl} +
\sum_{k\neq l}\, n_k (B_{kl} J_\nu + C_{kl}). \label{e:stateq}
\end{array}
\end{equation}
The collision rates $C_{kl}$ depend on the density and the collisional
rate coefficients of molecular hydrogen and other collision partners,
and on temperature through the detailed balance of the up- and
downward coefficients.  Eq.~(\ref{e:stateq}) can be easily solved
through matrix inversion for each position in the source provided the
radiation field $J_\nu$ is known.  However, $J_\nu$ contains
contributions by the CMB, dust and spectral lines, and since the
spectral line term depends on the level populations through
Eqs.~(\ref{e:di/ds1}), (\ref{e:jnu}) and (\ref{e:alpha}), the problem
must be solved iteratively.  Starting with an initial guess for the
level populations, $J_\nu$ is calculated, statistical equilibrium is
solved and new level populations are obtained; through the Monte Carlo
integration, the new populations yield a new value for $J_\nu$, after
which the populations are updated; etc., until the radiation field and
the populations have converged on a consistent solution.

When the physical conditions do not vary much over the model, an
approximate value of $J_\nu$ can be found from the local conditions
alone. This idea is the basis of the Large Velocity Gradient method,
the Sobolev method, microturbulence, or the escape probability
formalism
\cite[e.g.]{sobolev:book,goldreich:molclouds,leung:microt,dejong:cloudmodels}.
Also, in specific geometries, the integration over all solid angles
and along the full length of the line of sight of Eqs.
(\ref{e:di/ds2}) and (\ref{e:jbar}) can be greatly reduced, making the
problem tractable. This sort of technique has most application in
stellar and planetary atmospheres; the Eddington approximation is an
example.

However, in many astrophysical situations including the example of
\S~\ref{s:simplemodel}, such simplifications cannot be made, and
Eqs.~(\ref{e:di/ds2}) and~(\ref{e:jbar}) need to be fully solved to
get $J_\nu$. Compared to the relative ease with which statistical
equilibrium can be solved (Eq.~\ref{e:stateq}), obtaining $J_\nu$
becomes the central issue. Direct integration of Eqs.~(\ref{e:di/ds2})
and~(\ref{e:jbar}) with, e.g., Romberg's method, is infeasible for
realistic models, but based on a finite set of directions a good
approximation of $J_\nu$ can be obtained. The next two sections
describe two different methods to choose this set and construct
$J_\nu$ in this way.


\begin{figure}[t]
\begin{center}   
\psfig{figure=montecarlo.fig2,width=9cm,angle=-90}

\caption{(a) In the `traditional' formulation of the Monte Carlo
method for solving radiative transfer, the radiation field is
represented by a certain number of photon packages. Each of these
packages originates in a random position of the cloud, corresponding
to spontaneous emission, and travels in a random direction through the
cloud until it either escapes or is absorbed. To include the CMB
field, a separate set of packages is included, shown as dashed arrows,
that originate at the cloud's edge. The packages traversing a cell
during an iteration give $J_\nu$ in that cell.  (b) In our
implementation, an equivalent estimate of $J_\nu$ is found by choosing
a certain number of rays which enter the cell from infinity (or the
cloud's edge, using the CMB field as a boundary condition) from a
random direction and contribute to the radiation field at a random
point in the cell's volume. As \S \ref{s:acceleration} argues, this
formulation allows separation between the incident radiation field and
the locally produced radiation field, which accelerates convergence in
the presence of significant optical depth.\label{f:jbar}}
\end{center}
\end{figure}

\subsection{Constructing $J_\nu$ and the $\Lambda$-operator\label{s:jnu}}

For computational purposes, source models are divided into discrete
grid cells, each with constant properties (density, temperature,
molecular abundance, turbulent line width, etc.).  It is also assumed
that the molecular excitation can be represented by a single value in
each cell, which requires instantaneous spatial and velocity mixing of
the gas.  Appropriate source models have small enough cells that the
assumption of constant excitation is valid. The systematic velocity
field is the only quantity that is a vector field rather than a scalar
field, and in our code it is allowed to vary in a continuous way
within each cell. We divide the integration along a ray into subunits
within a cell to track the variation of the velocity projected on the
ray.

Such a gridded source model lends itself easily to the construction of
a finite set of integration paths to build up $J_\nu$. The average
radiation field can be thought of as the sum of the emission received
in cell $i$ from each of the other cells $j$ after propagation through
the intervening cells and weighted with the solid angle subtended by
each of these cells $j$ as seen from cell $i$.  The combination of
radiative transfer and statistical equilibrium can be written as
\begin{equation}
J_\nu = \Lambda\, [S_{ul}(J_\nu)]. \label{e:lambda}
\end{equation}
This equation states that the radiation field is given by an operator
$\Lambda$ acting on the source function $S_{ul}$, which depends on the
level populations and hence $J_\nu$ (Eqs. \ref{e:jnu}, \ref{e:alpha},
\ref{e:stateq}). Considering the narrow frequency interval around the
transition $u$--$l$, we have replaced $S_\nu$ by $S_{ul} \equiv
[j_{\nu_0}({\rm dust}) + \int j_\nu^{ul}({\rm gas}) d\nu] /
[\alpha_{\nu_0}({\rm dust}) + \int \alpha_\nu^{ul}({\rm gas})
d\nu]$. This corresponds to the assumption of instanteous
redistribution of excitation mentioned above.  In a gridded source
model, one can think of $\Lambda$ as a matrix describing how the
radiation field in cell $i$ depends on the excitation in all other
cells. The elements in the matrix then represent the radiative
coupling between cell pairs.

Eq.~(\ref{e:lambda}) can be solved iteratively, where an updated
value of $J_\nu$ is obtained by having $\Lambda$ operate on the
previous populations, $S_{ul}^\dag$,
\begin{equation}
J_\nu = \Lambda [S_{ul}^\dag(J_\nu)]. \label{e:lambda_iteration}
\end{equation}
Since $S_{ul}^\dag$ is already known, this only involves matrix
multiplication, compared to the much more expensive matrix inversion
required to solve Eq.~(\ref{e:lambda}). Because of this elegant
notation, iterative schemes for non-LTE excitation and radiative
transfer are commonly referred to as $\Lambda$-iteration, even if no
$\Lambda$-operator is ever actually constructed.
These methods share the use of the same set of rays throughout the
calculation, in contrast to Monte Carlo methods, which use random rays
as discussed in \S\S \ref{s:mc} and \ref{s:noise}.

Constructing the $\Lambda$-operator in multidimensional source models
is taxing on computer memory because of all the possible connections
to keep track of simultaneously. Techniques exist to reduce the number
of elements \cite{dullemond:radical}, but these are complex and
may require some fine-tuning for individual source
geometries. Alternatively, computer \emph{memory} can be exchanged for
computing \emph{time} by solving the problem serially, calculating the
radiation field in each of the cells due to the other cells one at a time.


\subsection{The Monte Carlo method\label{s:mc}}

One way of solving Eq.~(\ref{e:lambda_iteration}) is to directly sum
the contribution from all other cells to the radiation field in each
of the individual cells. This corresponds to replacing the integral in
Eq.~(\ref{e:jbar}) by a summation. With a judicially chosen fixed set
of directions or rays, as most $\Lambda$-iteration codes do, a good
approximation of $J_\nu$ can be found in this way \cite[e.g.]{phillips:phd}.
However, this procedure requires care, since the necessary angular
sampling depends, in principle, on characteristics of the excitation
solution of the problem at hand.

Since our aim is to construct a method that can be applied to many
different source models without too much fine-tuning, we adopt the
Monte Carlo approach to obtain $J_\nu$. Analogous to the Monte Carlo
method to solve the definite integral of a function [see chapter 7 of
Press et al. \cite*{numrep:ch7} for a discussion of Monte Carlo
integration, and further references], Eq.~(\ref{e:jbar}) can be
approximated by the summation over a \emph{random set} of
directions. This has the advantage that all directions are sampled to
sufficient detail: if too few directions are included, subsequent
realizations will give different estimates of $J_\nu$ (see
\S~\ref{s:noise} for further discussion of this issue).

Originally \cite{bernes:montec}, the Monte Carlo approach was phrased
in terms of randomly generated `photon packages', which are followed
as they travel through the source and which together approximate the
radiation field. Fig.~\ref{f:jbar} illustrates that a formulation in
terms of randomly chosen directions from each cell yields an
equivalent estimate of $J_\nu$. The only difference is the direction
of integration in Eq.~(\ref{e:di/ds1}). Where the former approach
follows the photons as they propagate through the cells, the latter
backtracks the paths of the photons incident on each cell.  As the
next section will discuss, this latter approach lends itself better to
improvements in its convergence characteristics. Treatment of
non-isotropic scattering is more complicated in this approach, and
since scattering is not important at the wavelengths of interest here,
$\ga 10 \mu$m, scattering is not included in the code.
Implementations of the Monte Carlo method more appropriate for
scattering are available in the literature
\cite{wood:scatter1,wood:scatter2,wolf:multidim}.


\subsection{Convergence and acceleration\label{s:acceleration}}

\begin{figure}[t]
 \begin{center}   
  \psfig{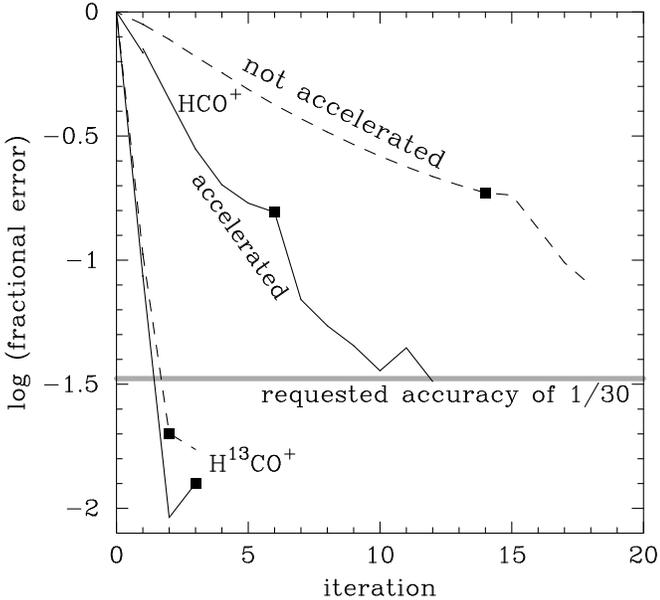}
  \caption{Evolution of the fractional error in the level populations
    of HCO$^+$ as function of iteration step with (`accelerated';
    solid line) and without (`not accelerated'; dashed line)
    separation of local and incident radiation field. For the
    optically thin H$^{13}$CO$^+$ molecule, both methods converge
    equally fast. The solid symbols indicate the iteration where the
    first stage of the calculation has converged (see \S
    \ref{s:noise}); after that random noise starts to dominate the
    fractional error, which is controlled by the increase in rays per
    cell. The source model is that described in \S
    \ref{s:simplemodel}. The `not accelerated' HCO$^+$ formally
  converged at iteration 18, because the difference with iteration 17
  became smaller than 1/30, even though the difference from the real
  solution exceeds that value. This illustrates that acceleration is
  not only computationally convenient, but may also essential for
  a correct solution.}
  \label{f:fracerr}
 \end{center}
\end{figure}

Besides estimating $J_\nu$, an important aspect of non-LTE radiative
transfer is convergence towards the correct solution in a reasonable
amount of time. Since the solution is not a priori known, convergence
is often gauged from the difference between subsequent iterative
solutions. This relies on the assumption that when $J_\nu$ and the
populations are far from their equilibrium solution, corrections in
each iteration are large. Large optical depth can be a major obstacle
to this behaviour: emission passing through an opaque cell will
rapidly lose all memory of its incident intensity and quickly tend
toward the local source function. The distance over which the
information about changes in excitation can travel is one mean free
path per iteration, so that the required number of iterations grows
$\propto \tau^2$ characteristic of random walk. This effect makes it
hard to determine if the process has converged.

Accelerated or Approximated Lambda Iteration \cite[ALI]{rybicki:ali1},
circumvents this problem by defining an approximate operator
$\Lambda^*$ such that
\begin{equation}
J_\nu = \left( \Lambda - \Lambda^* \right) [S_{ul}^\dag(J_\nu)] +
\Lambda^* [S_{ul}(J_\nu)]. \label{e:ali}
\end{equation}
An appropriate choice for $\Lambda^*$ is one which is easily
invertible and which steers rapidly toward convergence. This occurs if
$J_\nu$ is dominated by the second term on the right hand side of the
equation, where $\Lambda^*$ works on the \emph{current} source
function as opposed to the solution from the previous iteration. 

After several attempts \cite{scharmer:ali}, Olson, Auer, \& Buchler
\cite*{olson:ali} found that a good choice for $\Lambda^*$ is the
diagonal, or sometimes tri-diagonal, part of the full operator
$\Lambda$. This choice for $\Lambda^*$ describes the radiation field
generated locally by the material in each cell, and its direct
neighbours in the case of the tri-diagonal matrix. Eq.~(\ref{e:ali})
then gives $J_\nu$ as the sum of the field incident on each cell due
to the previous solution for the excitation \{$(\Lambda - \Lambda^*)
[S_{ul}^\dag]$\}, and a \emph{self-consistent} solution of the
\emph{local} excitation and radiation field \{$\Lambda^* [S_{ul}]$\}.
In opaque cells, the radiation field is close to the local source
function, and Eq.~(\ref{e:ali}) converges significantly faster than
Eq.~({\ref{e:lambda_iteration}); for optically thin cells, both
formalisms converge equally fast.

Formulating the Monte Carlo method in terms of randomly generated
photon packages traveling through the source does not easily permit
separation of the locally generated field and the incident field for
each cell. However, such a separation is possible when $J_\nu$ is
constructed by summation over a set of rays, which each start at a
random position within the cell and point in a random direction.  For
ray $i$, call the incident radiation on the cell $I_{0,i}$ and the
distance to the boundary of the cell $ds_i$.  The current level
populations translate this distance into an opacity $d\tau_i$, and
give the source function $S_{ul}$. The average radiation field from
$N$ rays then follows from Eqs.~(\ref{e:di/ds2}) and~(\ref{e:jnu}),
\begin{eqnarray}
J_\nu & = & \frac{1}{N}\, \sum_i I_{0,i}\, {\rm e}^{-\tau_i} +
            \frac{1}{N}\, \sum_i S_{ul}\, [1 -{\rm e}^{-\tau_i}]
\label{e:i0+snu} \\
      & = & J_\nu^{\rm external} + J_\nu^{\rm local}. 
\end{eqnarray}
Here, $S_\nu$ and $d\tau_i$ contain both line and continuum terms, and
$I_{0,i}$ includes the CMB. The radiation field is now the sum of the
external ($J_\nu^{\rm external}$) and internal ($J_\nu^{\rm local}$)
terms. Since the external term is evaluated using populations from the
previous Monte Carlo iteration (through $\tau_i$ and $S_{ul}$), this
scheme is akin to accelerated $\Lambda$-iteration. Within Monte Carlo
iterations, sub-iterations are used to find the locally
self-consistent solution of $S_{ul}$ and $\tau_i$ for given $J_\nu^{\rm
external}$.

The main computational cost of this strategy lies in following a
sufficient number of rays out of each cell through the source.
Iteration on Eq.~(\ref{e:i0+snu}) is relatively cheap and speeds up
convergence considerably in the presence of opaque cells.
Fig.~\ref{f:fracerr} illustrates this, by showing the evolution of the
fractional error of the solution of the simple problem posed in \S
\ref{s:simplemodel} for optically thick HCO$^+$ and thin
H$^{13}$CO$^+$ excitation (for a fixed set of directions -- see
below).

Population inversions require careful treatment in radiative transfer
codes, since the associated opacity is negative and the intensity
grows exponentially. In general, an equilibrium will be reached where
the increased radiation field quenches the maser. Since iterative
schemes solve the radiative transfer before deriving a new solution
for the excitation, the radiation field can grow too fast if
population inversions are present. Our code handles negative opacities
by limiting the intensity to a fixed maximum which is much larger than
any realistic field. Proper treatment requires that the grid is well
chosen, so that masing regions are subdivided into small cells where
the radiation field remains finite. Our code can deal with the small
population inversions that occur in many problems including the model
presented in \S \ref{s:simplemodel}. However, to model astrophysical
masers, specialized codes are required \cite[e.g.]{spaans:ohir}.


\subsection{The role of variance in Monte Carlo calculations\label{s:noise}}

Because the Monte Carlo method estimates $J_\nu$ from a randomly
chosen set of directions, the result has a variance, $\sigma$, which
depends on the number $N$ of included directions as $\sigma \propto
1/\sqrt{N}$. As explained above (\S~\ref{s:mc}), this variance is a
strength rather than a weakness of the Monte Carlo method. Since it is
not a priori known how many directions are required for a fiducial
estimate of $J_\nu$, this method automatically arrives at an
appropriate sampling by increasing $N$ until the variance drops below
a predefined value.

The variance of a solution is usually estimated from the largest
relative difference between subsequent iterations. In our
implementation (see appendix), the number $N$ of rays making up
$J_\nu$ in a particular cell is doubled each time the variance in that
cell exceeds a certain value; the variance is evaluated using the
largest relative difference between three subsequent solutions with
the same $N$.  This cell-specific scheme
ensures that the radiation field is sufficiently sampled everywhere,
and at the same time prevents oversampling of cells which are close to
LTE and/or weakly coupled to other regions.

The variance as estimated from the difference between subsequent
solutions only reflects the noise if random fluctuations dominate the
difference. There will be systematic differences between subsequent
solutions if these are still far from convergence. Therefore, many
Monte Carlo methods consist of two stages. In the first stage, a fixed
number of photons will yield a roughly converged solution; in the
second stage, the number of photons is increased until the noise
becomes sufficiently small.

In our implementation, this first stage consists of iterations with a
fixed number of directions making up $J_\nu$ in each cell, $N_0$,
which depends on the model. The directions are randomly distributed,
but in each iteration, the \emph{same} {\bf set} of random directions is
used by resetting the random number generator each iteration.  Without
random fluctuations in $J_\nu$, the difference between subsequent
solutions purely reflects the evolution toward convergence.  The first
stage is considered converged when this `noise' is a factor of
ten {\em smaller} than the user-specified level.

For a sufficiently large $N_0$ (typically a few hundred), the
excitation in each cell now is close to the final solution, except for
imperfections in the sampling of $J_\nu$. In the second stage, each
iteration uses a \emph{different} set of random directions to estimate
$J_\nu$: the random number generator is no longer reset. Based on the
resulting variance, the number of rays in each cell is doubled each
iteration, until the noise on the level populations in each cell is
below a given value. If $N_0$ was initially insufficient, the variance
will contain a significant contribution from systematic differences
between iterations. Even though this will slow down the code by
artificially increasing the number of rays in these cells as the code
over-compensates the variance, ultimately the code will still converge
to the correct solution.

The separation between local and incident radiation fields in our
method (\S \ref{s:acceleration}) keeps the system responsive to
changes even in the presence of significant optical depth. This
accelerates the convergence, but also increases the noise level. The
literature mentions several methods to reduce the noise of Monte Carlo
methods, e.g., with a reference field \cite{bernes:montec,choi:b335mc}
or quasi-random instead of pseudo-random numbers \cite{juvela:clumpy}.
These schemes are useful when assumptions about the solution are
available, but may slow down convergence if the initial guess is far
off.  Since the `first stage' described above and the presence of
noise prevents Monte Carlo methods from `false convergence', we have
not included any noise reduction techniques in our code, to keep it as
widely applicable as possible.


\subsection{Implementation and performance characteristics}
\label{sec:impl}

\begin{figure*}[t]
\begin{center}
\psfig{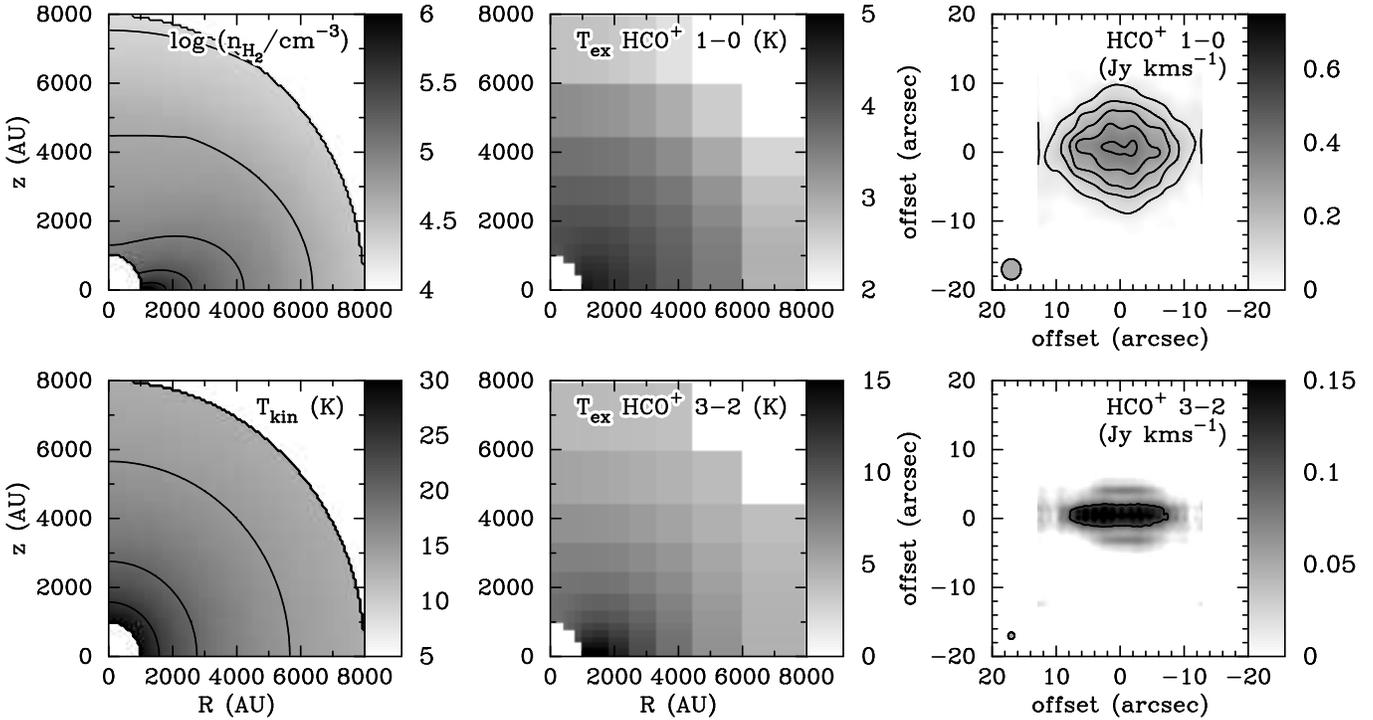}
\caption{Left: Density distribution (top) and temperature distribution
(bottom) of the collapse model including rotation of \S
\ref{s:flat}. Middle: Resulting excitation temperature of HCO$^+$ 1--0
and 3--2. The adopted grid is visible in these panels, with small
cells at the center were the density is large, and larger cells in the
outer regions of the object. Right: Images of integrated intensity of
HCO$^+$ 1--0 and 3--2 for an edge-on source orientation, after
sampling on spatial frequencies corresponding to interferometric
baselines between 15 and 300 m and subsequent image
reconstruction. This results in synthesized beam sizes of $3''$ and
$1''$ for the 1--0 and 3--2 lines, respectively, as indicated in the
lower left corner of the panels.
\label{f:flat}}
\end{center}
\end{figure*}

Appendix~\ref{a:code} describes the structure of the program in
detail, and provides a reference to a web site where we have made the
source code of its spherically symmetric version publicly
available. To test its performance, Appendix~\ref{a:tests} compares
results obtained with our code to those of other codes.

The main part of the program deals with calculating the excitation
through the source model. In a separate part, comparison to
observations proceeds by integrating Eq.~(\ref{e:di/ds1}) on a grid of
lines of sight for a source at a specified inclination angle and
distance. The resulting maps of the sky brightness may be convolved
with a beam pattern for comparison with single-dish data, or Fourier
transformed for comparison with interferometer data.

Appendix~\ref{a:tests} describes tests of the validity of the results
of the program.  We have also tested up to what optical depth the
program can be used, and found that this depends on source model.
These tests were done on a Sun Ultrasparc~2 computer with 256~Mb
internal memory and a 296~MHz processor. For a simple, homogeneous
HCO$^+$ model with $n=10^4$~cm$^{-3}$ and $T=30$~K, the code produces
accurate results within an hour for values of $N$(HCO$^+$) up to $\sim
10^{17}$~cm$^{-2}$, corresponding to $\tau \sim 20,000$ in the lowest
four rotational transitions. Higher$-J$ lines are less optically thick
under these physical conditions. For such opaque models, `local'
approximations fail badly, because the excitation drops sharply at the
edge of the model (Bernes 1979; Appendix~\ref{a:code}). 

For a power-law, Shu-type model, performance is somewhat slower. The
dense and warm region fills only a small volume, while its radiation
has a significant influence on the excitation further out, and
modeling this effect requires a large number of rays. We have used the
specific model from the Leiden workshop on radiative transfer
(Appendix \ref{a:tests}) for various values of the HCO$^+$ abundance.
Within a few hours, accurate results are obtained for values of
HCO$^+$/H$_2$ up to $10^{-6}$, corresponding to $\tau = 600 - 2000$ in
the lowest four lines.

These test cases should bracket the range of one-dimensional models of
interest. For two-dimensional models, the limitations of present-day
hardware are much more prohibitive. As a test case, we have used the
flattened infalling envelope model from \S~\ref{s:flat} for various
HCO$^+$ abundances. Within 24~hours, the above machine provides a
converged solution for HCO$^+$/H$_2$ up to $10^{-8}$, corresponding to
a maximum optical depth of $\sim 100$. Realistic models often have
higher opacities, and call for the use of parallel computers. However,
as faster computers are rapidly becoming available, we expect that
these limitations will become less relevant in the near future. For
both one- and two-dimensional models, the second, ray-tracing part of
the code to create maps of the sky brightness takes only a fraction of
the computer resources of the first part.

\subsection{Alternative accelerators\label{sec:ng}}

Another method to tackle slow convergence in the presence of large
opacities is \emph{core saturation} \cite{rybicki:coresat,doty:models},
where photons in the optically thick line center are replaced by the
source function and no longer followed, while photons in the still
optically thin line wings which carry most information are more
accurately followed.  This scheme has been implemented in a Monte
Carlo program by Hartstein \& Liseau \cite*{hartstein:mch2o}, but
involves a choice where to separate the line core from the line wings.
Since the effectiveness of the method depend on this choice, we have
not included core saturation in our program.

A completely different approach to accelerating convergence is to
extrapolate the behaviour of the populations from the last few
iterations. This so-called Ng acceleration \cite{ng:acceleration} is not
implemented in our code, because extrapolating from an inherently
noisy estimate may be dangerous.


\section{Astrophysically relevant examples\label{s:examples}}

A first example of the application of our code has already been given
in \S\S \ref{s:simplemodel} and \ref{s:acceleration}. This model is a
spherically symmetric (one-dimensional) model; many authors have
already illustrated the capability of Monte Carlo and other methods in
solving one-dimensional problems
\cite[e.g.]{bernes:montec,zhou:rotation,choi:b335mc}. This
section presents a number of astrophysically relevant examples, again
drawn from star formation studies, to illustrate the distinguishing
properties of our code -- in addition to accelerated convergence: the
capability to calculate axisymmetric source models with a wide range
of spatial scales, and the effects of dust continuum on radiative
transfer and excitation.

The models presented in the following sections all include continuum
radiation fields from dust. For these star-forming regions, we have
chosen the model of Ossenkopf \& Henning \cite*{ossenkopf:kappa} for
the dust emissivity, which includes grain growth for a period of
$10^5$ yr at an ambient density of $10^6$ cm$^{-3}$ and thin ice
mantles. Except for the calculations in \S \ref{s:himass} where we
specifically examine the effect of dust on the excitation including
infrared transitions, only (sub) millimetre transitions were included
in the excitation calculations which are not significantly influenced
by the relatively weak continuum field.


\subsection{A young stellar object with rotation\label{s:flat}}

\begin{figure*}[t]
\begin{center}
\psfig{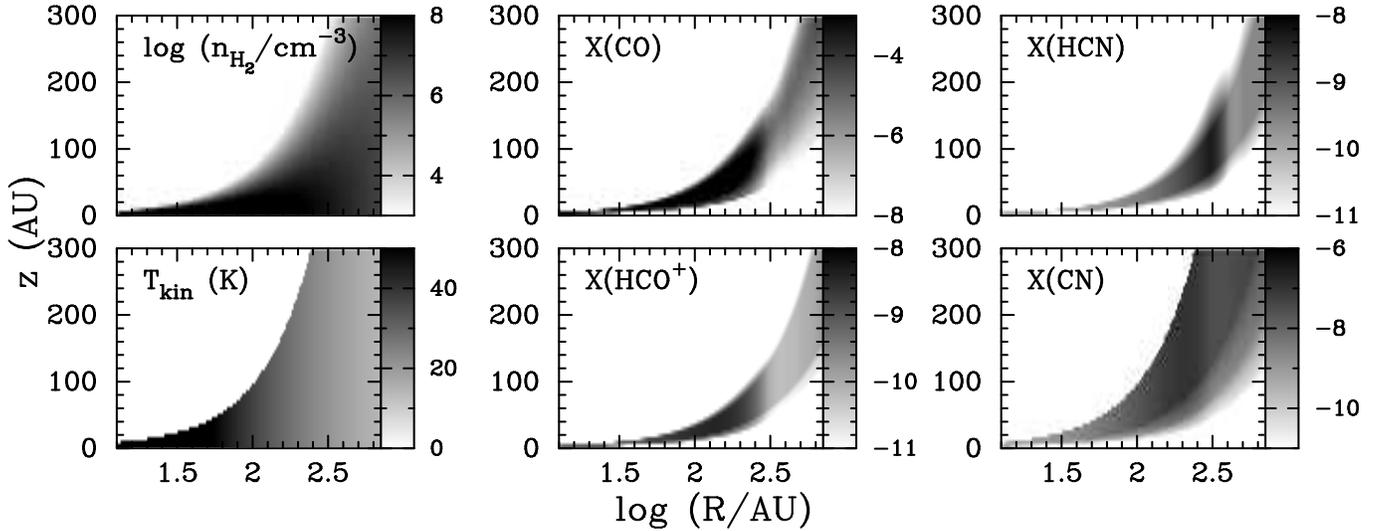}
\caption{The density, temperature, and molecular abundance
distribution of the circumstellar disk model described in \S
\ref{s:disk}. Density and abundances are plotted on logarithmic
scales; the temperature is plotted on a linear scale. The
`superheated' layer of the disk model of Chiang \& Goldreich is not
included in our model because only a very small amount of dust and
gas is present in this layer. Its `backwarming' effect on the disk
interior is included, however.
\label{f:disk_model}}
\end{center}
\end{figure*}

\begin{figure*}[t]
\begin{center}
\psfig{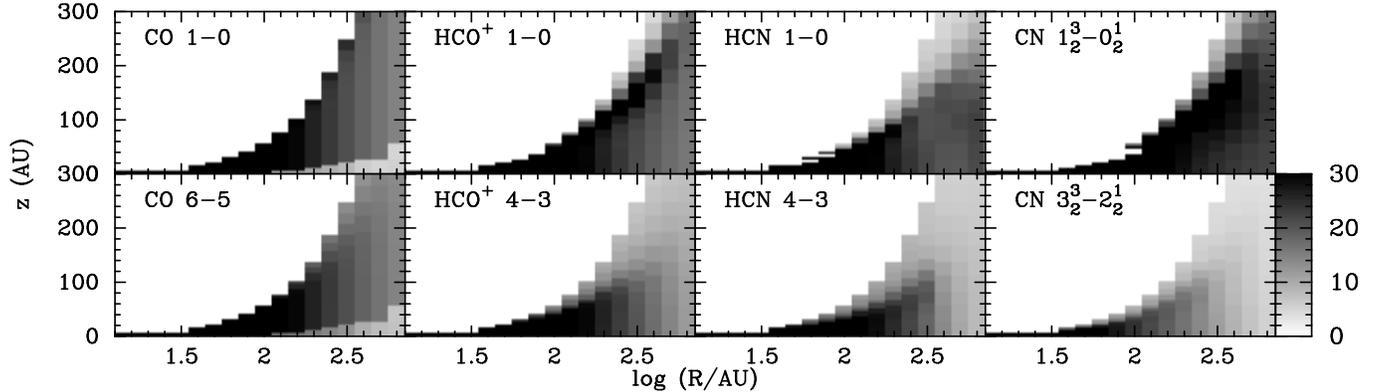}
\caption{The distribution of excitation temperatures in K for the $J$=1--0
and selected submillimetre lines of CO, HCO$^+$, HCN, and CN. In all
panels the grey scale ranges between 0 K and 30 K on a linear scale.
The adopted grid, exponentially spaced in radius and linearly in
height, is reflected in the excitation distribution.\label{f:disk_tex}}
\end{center}
\end{figure*}

Observations of nearby young stellar objects often show flattened
structures rather than the spherical symmetry of models like that of
Shu \cite*{shu:selfsim}, presumably caused by ordered magnetic fields
and/or rotation \cite[e.g.]{mrh:taurus2}.  These mechanisms probably
influence the accretion behaviour, and may give rise to a rotationally
supported circumstellar disk. To test these ideas against
observations, cylindrically symmetric source models need to be
considered.  This section examines a model of protostellar collapse
that includes flattening due to rotation following the description of
Terebey, Shu, \& Cassen \cite*{terebey:tsc}, and its appearance in
aperture synthesis maps.

The model of Terebey et al.\ \cite*{terebey:tsc} treats rotation as a
small perturbation to the solution of Shu \cite*{shu:selfsim} for a
collapsing envelope, joined smoothly to the description of a
circumstellar disk by Cassen \& Moosman \cite*{cassen:disks}. In
addition to the sound speed and age, which we take identical to the
values of \S \ref{s:simplemodel} of $a=0.24$ km~s$^{-1}$ and
$t=1\times 10^5$ yr, this model is parameterized by a rotation rate
$\Omega$. This gives rise to a centrifugal radius $R_c \equiv a m_0^3
t^3 \Omega^2 / 16$, within which the infalling material accretes onto
a thin disk. Here, $m_0=0.975$ is a numerical constant. We choose
$\Omega=5.9\times 10^{-13}$ s$^{-1}$, so that $R_c = 800$ AU. We
assume that inside $R_c$ the material accretes onto a thin disk, and
that most molecules rapidly freeze out onto dust grains (cf.\ \S
\ref{s:disk}). Effectively, we assume the region within $R_c$ to be
empty for this calculation. Fig.~\ref{f:flat} (top left) shows the
adopted density structure. All other parameters are similar to the
model of \S \ref{s:simplemodel}.

To follow the power-law behaviour of the density in the model, a total
of $15 \times 15$ cells are spaced exponentially in the $R$ and $z$
directions.  To reach a final signal-to-noise ratio of 10, with 300
rays initially making up the radiation field in each of the cells, the
Monte Carlo code requires 5 hours CPU time on a UltraSparc 10 to
converge on the HCO$^+$ solution. For comparison, the optically thin
and more readily thermalized $^{13}$CO problem takes only 10
minutes. The resulting excitation temperature distribution
(Fig.~\ref{f:flat}; middle panels) does not deviate much from that of
the spherically symmetric model of \S \ref{s:simplemodel}, apart from
the flattened distribution of the material at the center: rotation is
only a small perturbation on the overall source structure. As a
result, the appearance is mostly unaffected in single-dish
observations which do not resolve scales comparable to $R_c$ where
flattening is important.

Millimetre interferometers, on the other hand, can resolve these
scales at the distances of the nearest star-forming regions, $\sim
140$ pc. Fig.~\ref{f:flat} (right panels) shows the integrated
emission in HCO$^+$ $J$=1--0 and 3--2 after sampling at the same
spatial scales as real interferometer observations and subsequent
image reconstruction. Delays between 100--1000 ns were used,
corresponding to angular scales of $2''$--$40''$ for the 1--0
line and $0{\farcs}6$--$13''$ for 3--2.  Hence, emission on scales
$\ga 6000$~AU ($\ga 2000$ AU at 3--2) is filtered out. This results in
a reconstructed (`cleaned') image that is dominated by the central,
flattened regions of the envelope when the object is seen edge-on.
Aperture synthesis observations of embedded protostars in Taurus show
similar structures \cite[e.g.]{ohashi:l1527,mrh:taurus2}.


\subsection{A circumstellar disk\label{s:disk}}

\begin{figure*}[t]
\begin{center}
\psfig{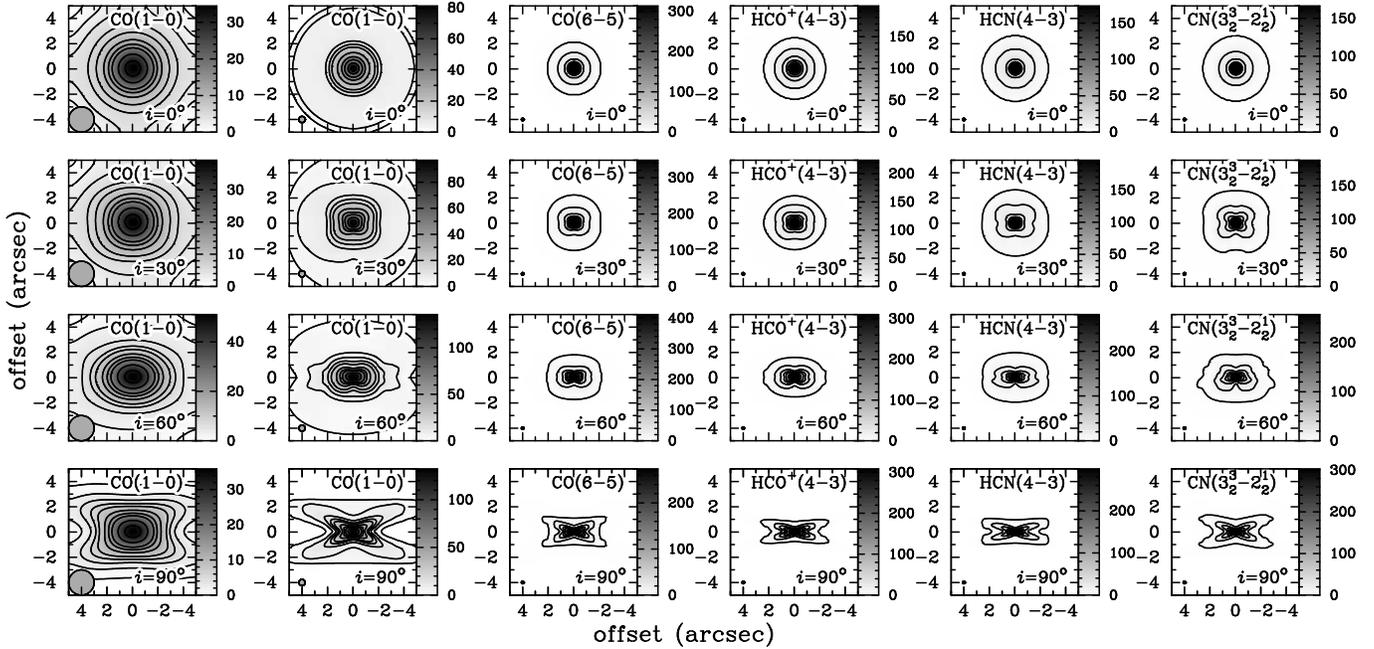}
\caption{Appearance of integrated intensity of selected lines in the
  circumstellar disk model at resolutions attainable with current and
  planned interferometric facilities. The first three columns from the
  left show CO 1--0 at $2''$ resoltuion, CO 1--0 at $0{\farcs}5$ resolution, 
  and CO 6--5 at $0{\farcs}2$ resolution. A distance of 140 pc is
  adopted for the source, and inclination increases from $0^\circ$ to
  $90^\circ$ from top to bottom. The last three columns from the left
  show the emission in HCO$^+$ 4--3, HCN 4--3, and CN $3_{3/2}\to
  2_{1/2}$ at $0{\farcs}2$ resolution. All panels are shown with a
  linear grey scale ranging from 0 to the image maximum in K~km~s$^{-1}$.
  Contours are drawn at 1\%, 5\%, 10\%, 15\%, 20\%, 30\%, 40\%,
  50\%, 70\%, and 90\% of maximum.
  \label{f:disk_emission}}
\end{center}
\end{figure*}

Planetary systems form from the disks that surround many young stars
\cite{beckwith:crete2,ppiv}.  Observational characterization of these
disks is of prime importance to increase our understanding of the
processes that shape planetary systems. Here, we present simulations
of molecular line observations of a circumstellar disk around a
T~Tauri star as obtained with current and planned
millimetre-interferometric facilities.

The physical structure of the model disk is that of a passive
accretion disk in vertical hydrostatic equilibrium as described by
Chiang \& Goldreich \cite*{chiang:disksed}. This description includes
the effect of `backwarming' of the disk by thermal radiation of a
thin, flared surface layer that intercepts the stellar light.  The
total amount of material in the superheated surface layer is too small
to be detectable, but the overall effect of increased mid-plane
temperature is significant. The effective temperature of the central
star is 4000~K and its luminosity is 1.5 L$_\odot$. The outer radius
of the disk is 700~AU, with a total mass of 0.04 M$_\odot$.

The chemical structure of the disk follows Aikawa \& Herbst
\cite*{aikawa:chem2d}, who describe the radial and vertical
composition of a flared disk. Freezing out of molecules onto dust
grains is one of the dominant processes influencing the gas-phase
composition in disks, and strongly depends on temperature and density.
In the dense and cold midplane, many molecules will be depleted onto
grains. However, close to the star where temperatures are raised, and
away from the midplane where densities are lower and depletion time
scales longer, gas-phase abundances will be significant. In addition,
ultraviolet radiation and X-rays may penetrate the upper layers of the
disk, photodissociating molecules and increasing the abundance of
dissociation products like CN. Fig.~\ref{f:disk_model} shows the
distribution of the density, temperature, and abundances of CO,
HCO$^+$, HCN and CN in the adopted model. We have used the results
presented in Aikawa \& Herbst \cite*[their Figs. 6 and 7; high
ionization case]{aikawa:chem2d}, and parameterized the abundances as
function of scale height.

For the Monte Carlo calculations, a gridding is adopted that follows
the radial power-law density profile in 14 exponentially distributed
cells and the vertical Gaussian profile in 13 cells linearly
distributed over 3 scale heights. Convergence to a signal-to-noise
ratio of 10 requires approximately 6 hours CPU time per model on an
UltraSparc 10 workstation, starting with 100 rays per cell and
limiting the spatial dynamic range to 36 (i.e., the smallest cell
measures 10 AU on the side). The resulting excitation and emission
depends on the competing effects of increased abundance and decreased
density with distance from the midplane.
Fig.~\ref{f:disk_tex} shows the excitation temperature of selected
transitions and molecules. Fig.~\ref{f:disk_emission}
shows a number of representative simulated observations, at
resolutions of $2''$, $0{\farcs}5$, and $0{\farcs}2$ attainable with
current and planned (sub) millimetre interferometers.  Van Zadelhoff
et al. (in prep.) present a simpler analysis of similar models.


\subsection{A high mass young stellar object\label{s:himass}}

\begin{figure}[t]
  \begin{center}
    \psfig{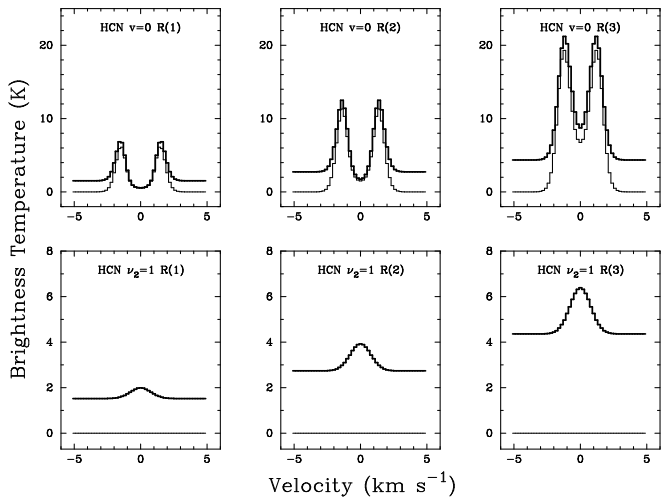}
    \caption{Submillimetre emission lines in a $1''$ beam of HCN in
    the vibrational
      ground state (top panels) and first excited state (bottom
      panels). Thin line: only collisional excitation; thick line:
      model with pumping through the $\nu_2$ band at $14$ $\mu$m.}
    \label{f:dust_hcn}
  \end{center}
\end{figure}

The above examples referred to the formation of stars with masses of
up to few M$_\odot$ and luminosities $\la 100$ L$_\odot$. Stars of
higher mass spend their first $\sim 10^4$ yr in envelopes of $\sim
100$ M$_\odot$. With their luminosities of $10^4$--$10^5$ L$_\odot$,
these stars heat significant parts of their envelopes to several
hundred K, shifting the peak of the Planck function to the wavelengths
of the vibrational transitions of many molecules. Stars of lower mass
and luminosity only heat small regions to a few hundred K, and the
impact on the excitation is correspondingly smaller.  As an example,
Figure~\ref{f:dust_hcn} shows two models of the HCN submillimetre line
emission, with and without pumping through the bending (``$\nu_2$'')
mode at $14.02$~$\mu$m. For computational convenience, only energy
levels up to $J$=10 within the first vibrationally excited and ground
states have been included, including l-type doubling in the excited
state. This doubling occurs due to the two possible orientations of
the rotational and vibrational motions with respect to each
other. Collisional rate coefficients between rotational levels are
from Green (1994, priv.\ comm., see
http://www.giss.nasa.gov/data/mcrates); within vibrational levels,
they were set to $10^{-12}$~cm$^3$s$^{-1}$. The source model is that
of the young high-mass star GL~2136 by van der Tak et al.\
\cite*{tak:massive}. Based on its luminosity of $7\times 10^4$
L$_\odot$ and dust mass of $\sim 100$ M$_\odot$, the star has heated a
region of radius $\sim 3000$~AU to $\ga 100$~K, making pumping through
the $14$ $\mu$m HCN bending mode important. We have assumed that
$T_{\rm gas}=T_{\rm dust}$, as is true for high density regions. The
dust emissivity, from Ossenkopf \& Henning \cite*{ossenkopf:kappa}, is
the same as in the previous sections. As seen in
Fig.~\ref{f:dust_hcn}, the effect of dust is especially strong for the
rotational lines within the $\nu_2$=1 band, which occur at frequencies
slightly offset from the ground state transitions. These lines have
indeed been detected towards similar objects
\cite[e.g.]{ziurys:vibhcn}.

The shells around evolved stars is another area where inclusion of
infrared pumping by dust is essential to understand the rotational
line emission \cite[e.g.]{ryde:15194co}. Many molecules that are
commonly observed through rotational lines at millimetre wavelengths
have ro-vibrational bands in the mid-infrared, and can be pumped by
warm dust. In a few cases, pumping through \emph{rotational} lines at
far-infrared wavelengths is important as well, for example CS
\cite{hauschildt:cs10-9} and all hydrides, most notably water
\cite{hartstein:mch2o}.


\section{Conclusion\label{s:conclusion}}

This paper describes a computer code to solve the coupled problem of
radiative transfer and molecular excitation for spherically and
cylindrically symmetric source geometries. It is based on the Monte
Carlo method, but incorporates elements from accelerated lambda
iteration which greatly improve convergence in the presence of
significant optical depth. In particular, the code separates
excitation due to the local environment from the response to the
radiation field after propagation through the source. This approach
combines the flexibility of a Monte Carlo method with the reliability
of accelerated lambda iteration. We expect our code to be a valuable
tool in the interpretation of the wealth of data that current and
future instruments operating from the millimetre to the infrared will
yield. A number of examples (\S \ref{s:examples}) already illustrates
the applications to problems in star formation studies.

\begin{acknowledgements}
  The authors wish to thank Ewine van Dishoeck for many stimulating
  discussion on the topic of this paper, her careful reading of the
  manuscript, and making available a data base with molecular
  parameters for this code; David Jansen for maintaining this data
  base; Lee Mundy for assistance with the construction of an earlier
  version of the part of the code calculating the sky brightness
  distributions; Marco Spaans for discussions and hospitality at the
  Astronomy Department of the Johns Hopkins University where an
  initial version of the code was written; Minho Choi and Neal Evans
  for the use of their Monte Carlo program for testing purposes; the
  organizers (Gerd-Jan van Zadelhoff, Kees Dullemond and Jeremy Yates)
  and participants of the May 1999 workshop on Radiative Transfer in
  Molecular Lines at the Lorentz Center of Leiden University; and
  George Rybicki for suggesting that iterating on the level
  populations may be of benefit in a Monte Carlo code.  The research
  of M. R. H. is supported by the Miller Institute for Basic Research
  in Science; that of FvdT by NWO grant 614-41-003.
\end{acknowledgements}


\appendix

\section{The code\label{a:code}}

\begin{figure}[t]
  \begin{center}
   \psfig{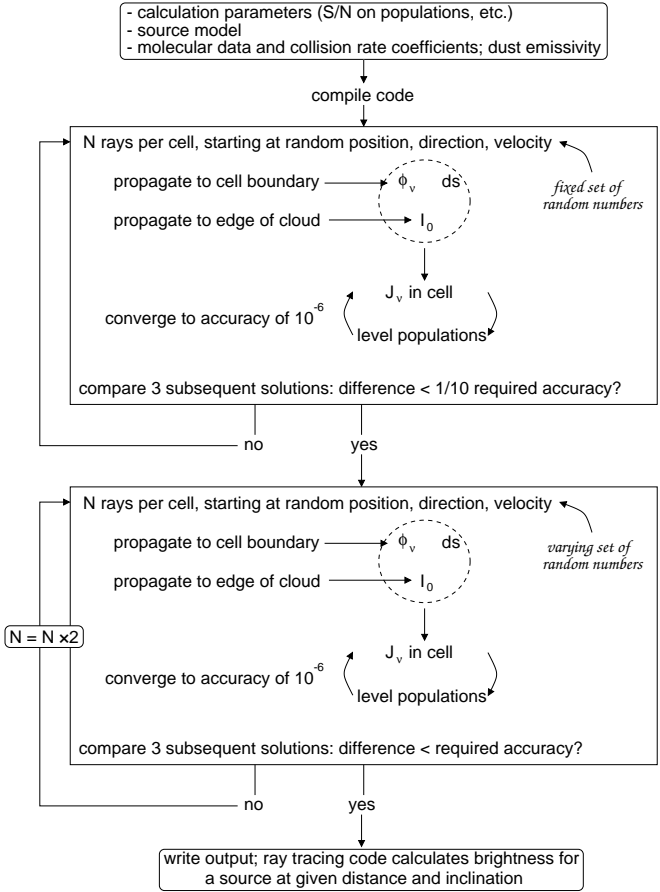}
    \caption{Schematic outline of steps involved in our Monte Carlo
      calculations.}
    \label{f:flowchart}
  \end{center}
\end{figure}

Our code is applicable to a wide range of astrophysical problems
involving (sub) millimetre and infrared ($\ga 10$ $\mu$m)
spectral-line and continuum observations. The one-dimensional
(spherically symmetrical) version of our code is publicly available
for all interested researchers via
http://astro.berkeley.edu/$\sim$michiel. The two-dimensional
(cylindrically symmetric) code is available on a collaborative basis
through the authors (see the same website for contact information).
This appendix gives a concise description of the implementation of the
accelerated Monte Carlo method.

Fig.~\ref{f:flowchart} gives an overview of the structure of our code,
which consists of two parts. The first runs the Monte Carlo simulation
solving the radiative transfer and molecular excitation. The second
part uses this solution to calculate the emission that would be
observed from this source above the atmosphere and with perfect
spatial and velocity resolution, given a source distance and, for
cylindrically symmetric models, inclination. This latter part can also
be used to calculate the continuum radiation emitted by the
source. Its output format is that used by the MIRIAD package
[Multichannel Image Reconstruction, Image Analysis, and Display; Sault
et al.\ \cite*{sault:miriad}]. This package, designed to analyse
interferometer interferometric spectral line data, includes many
processing options such as convolution with a single-dish beam and
modeling of aperture-synthesis visibilities, as well as a wide variety
of imaging capabilities. MIRIAD also allows easy conversion to the
ubiquitous FITS format and portability to other software packages.

Both parts of the code are controlled by UNIX C-shell scripts that
extract information from the provided input and compile an
executable code. In this way, the size of several arrays containing
the source model, the collisional rate coefficients, etc., can be
adjusted to the required size, minimizing memory requirements. The
source code is written in FORTRAN-77.

Following the flow chart of Fig.~\ref{f:flowchart}, the following
steps describe the Monte Carlo part of the code in more detail.

\begin{enumerate}
  
\item{The code starts by reading a list of keywords, detailing the
    required signal-to-noise ratio on the level populations, the
    initial number of photons in each cell ($N_0$), and pointers to the
    source model, the systematic velocity field (if any), the
    description of the dust emissivity, and the molecular energy
    levels and collisional rate coefficients. The velocity field can
    be defined simply through the source model with each grid cell
    moving at a constant speed, or it can be a constantly varying
    function over each cell. The source model can be a series of
    concentric shells covering a region from the origin to a maximum
    radius, or a series of stacked cylinders fully covering a region
    out to a maximum radius and height.
    
    Collisional rate coefficients are available for many
    astrophysically interesting species and common collision partners
    such as H$_2$ in the $J=0$ and in the $J=1$ levels, e$^-$, and
    He. Our code currently allows for two simultaneous collision
    partners, e.g., H$_2$ and e$^-$, each with its own density and
    temperature. For molecular ions such as HCO$^+$, excitation due to
    collisions with electrons can be significant compared to
    collisions with H$_2$ at fractional ionization ($\ga
    10^{-5}$). Often, listed rate coefficients are equivalent rates
    per H$_2$ molecule including contributions from He at cosmic
    abundance. The results of our code, and any non-LTE calculation,
    sensitively depend on the quality of the rate coefficients.
    Recently, Black \cite*{black:iau197} discussed the need for good
    rate coefficients and the effects of other implicit assumptions of
    radiative transfer codes.}
  
\item{In the first stage of the calculation, the radiation field is
    based on $N_0$ rays per cell, each starting at a random position
    equally distributed over the cell volume, pointing in a random
    direction, and at a random frequency within 4.3 times the local
    line width around the local systematic velocity vector. The value
    of 4.3 corresponds to the width where the line profile has dropped
    to less than 1\% of its peak. In this stage, in each iteration
    the \emph{same} series of random numbers is used, so that there
    are no random fluctuations in the coverage of the radiation
    field.}

\item{For each ray, the distance $ds$ from the ray's origin to the
    nearest boundary of the cell along its randomly chosen direction
    is calculated.  The incident radiation $I_0$ along the ray then
    follows from integrating Eq.~(\ref{e:di/ds1}) in a stepwise manner
    from cell edge to cell edge, attenuating the contribution from
    each cell by all intervening cells, with the 
    cosmic microwave background as a boundary condition. The only
    quantity that changes when stepping through a cell is the
    direction, and possibly the magnitude, of the systematic velocity
    vector, which enters Eq.~\ref{e:di/ds1} through the line profile
    function $\phi(\nu)$.  Changes of $\phi(\nu)$ within cells are
    tracked by subdividing the integration into small steps as needed.} 

\item{Armed with the set of $(ds,\phi(\nu),I_0)$ for each ray, the
radiation field $J_\nu$ in the cell follows from
Eq.~(\ref{e:i0+snu}). A consistent solution of this equation and the
level populations (Eq.~\ref{e:stateq}) quickly follows from iteration
to a relative accuracy of $10^{-6}$ in the populations. Limitations on
masering are discussed in Sect.~\ref{s:acceleration}.}

\item{The first stage of the code repeats items 2--4 until the largest
    relative fractional difference between the populations in all
    cells of three subsequent solutions is ten times better than
    ultimately required. Since the angular sampling is the same in
    each iteration, these differences are free of random noise but
    might not adequately sample all directions and frequencies.}

\item{The second stage of the code proceeds along similar lines as the
first stage, but with a \emph{different} set of random numbers in each
iteration. The only other difference is, that each time the maximum
fractional error in the populations in a cell exceeds the requested
accuracy, the number of rays in that cell is doubled. This stage lasts
until all cells comply with the required accuracy, after which the
solutions are written out to a file.}
 
\end{enumerate}

The second part of the program calculates the emission distribution on
the sky for a given source distance and inclination by simple ray
tracing. The output from the Monte Carlo code forms the input for this
ray-tracing code. Since it uses much of the same code as the Monte
Carlo part, geometry and radiative transfer being the same, it is not
further discussed here.

\section{Comparison with other codes\label{a:tests}}

This section describes two cases to test our code against
well-documented calculations with Monte Carlo codes from the
literature. For further tests, we refer the reader to the web-page
collecting a number of standard test cases, which has resulted from
the 1999 workshop on Radiative Transfer in Molecular Lines at the
Lorentz Center of Leiden University
(http://www.strw.leidenuniv.nl/$\sim$radtrans).

\subsection{Bernes' CO cloud}

  \begin{figure}[t] \begin{center}
    \psfig{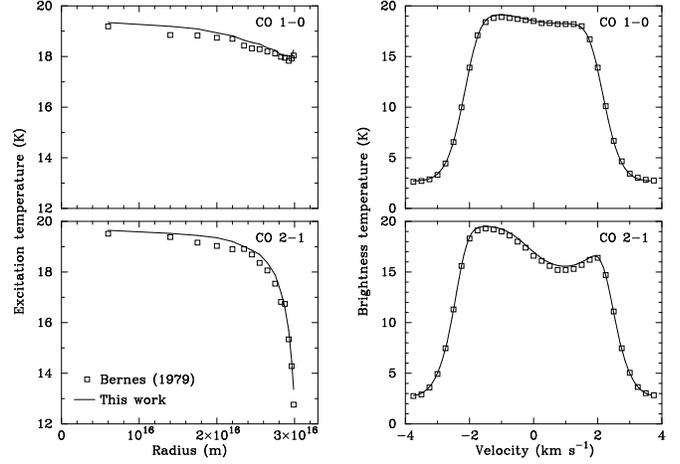}
    \caption{Excitation temperature of CO $J=1\to0$ and $2\to1$ as a
    function of radius, and integrated line profiles. Open symbols are
    results by Bernes (1979), the solid lines are our calculations.}

      \label{f:bernes}
    \end{center}
  \end{figure}
  
  In his seminal paper on Monte Carlo methods for radiative transfer
  and molecular excitation, Bernes \cite*{bernes:montec} presents a
  constant-density, constant-temperature, optically thick cloud model.
  The density of the cloud, $n_{\rm H_2} = 2000$ cm$^{-3}$, is below
  the critical density of the CO transitions, and the excitation is
  dominated by radiative trapping. The excitation temperatures of the
  CO transitions drop off rapidly in the outer regions of the cloud.
  This necessitates fine sampling of these regions.
  Figure~\ref{f:bernes} shows that our code reproduces the original
  results within the accuracy of our and Bernes' calculations.  This
  simple model forms a critical test for the code's ability to
  correctly handle excitation by radiative trapping. The total run
  time for the model was approximately 5 minutes on a UltraSparc 10
  workstation, using the same collisional rate coefficients as Bernes
  (Table~\ref{t:rates}).

\subsection{Model for B~335 by Choi et al.\label{a:choi}}

\begin{figure}[t] 
\begin{center}
\psfig{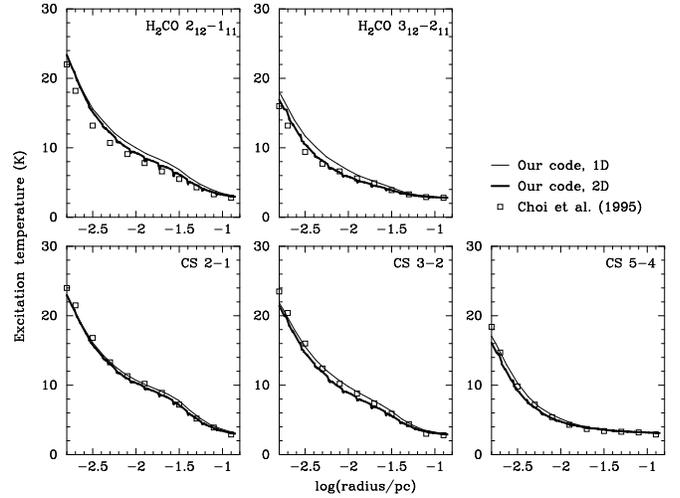}
 \caption{Excitation temperature of selected CS and H$_2$CO lines versus
radius. Open symbols are results by Choi et al.\ (1995), the thin
 solid lines are the results from our spherically symmetric code. The
 thick solid lines show the results from our cylindrically symmetric
 code. \label{f:choi} }
\end{center}
\end{figure}

Another critical element of any radiative transfer code is its ability
to correctly deal with systematic velocity fields. The inside-out
collapse model as outlined in \S \ref{s:simplemodel} is well suited
for such a test, because of its wide range in velocities from zero to
many times the turbulent line width combined with significant optical
depth. As a test case, we calculate the populations and the emergent
spectrum of several CS and H$_2$CO lines, following the model for the
young stellar object B~335 of Choi et al.\ \cite*{choi:b335mc}. This
model is similar to that of \S~\ref{s:simplemodel}, with $a=0.23$
km~s$^{-1}$ and $t=1.3\times 10^5$ yr. The turbulent line width is
0.12 km~s$^{-1}$.  Only the temperature structure is different from
\S~\ref{s:simplemodel}: Choi et al.\ use continuum observations to
constrain the temperature distribution, which we follow as closely as
possible from their Fig.~3.

Fig.~\ref{f:choi} compares the resulting excitation temperature
distribution with the results of Choi et al. The agreement is very
good for CS, where we used the same collisional rate coefficients
(Table~\ref{t:rates}). For H$_2$CO the agreement is less favourable,
but we were unable to use the exact same rate coefficients. Simple
tests show that the deviation is comparable to what can be expected
from the difference in the molecular data. This variation
corresponds to a 10\% difference in the emergent line profiles.

Fig.~\ref{f:choi} also plots the excitation temperatures obtained for
the same model but using the cylindrically symmetric code. Both codes
clearly give consistent answers; the small `wiggles' in the excitation
temperatures as function of radius in the output of the cylindrically
symmetric calculation can be attributed to geometrical defects when
trying to fit a sphere in a series of stacked cylinders.




\end{document}